\documentclass{iopart}
\usepackage{graphicx}  
\usepackage{latexsym}  

\jl{6}        
\eqnobysec    

\def\beq{\begin{equation}}
\def\eeq{\end{equation}}

\def\rmd{{\rm d}}

\def\rightcontract{\mathop{\hbox{\vrule width0.5pt height6pt%
  \vrule height0.5pt width6pt}}}

\begin{document}

\title[Spinning test particles and clock effect in Schwarzschild spacetime]
{Spinning test particles and clock effect in Schwarzschild spacetime}

\author{
Donato Bini$^* {}^\S{}^\P$,
Fernando de Felice$^\dagger$ 
and 
Andrea Geralico$^\ddag {}^\S$}
\address{
  ${}^*$\
Istituto per le Applicazioni del Calcolo ``M. Picone'', CNR I-00161 Rome, Italy
}
\address{
  ${}^\S$\
  International Center for Relativistic Astrophysics,
  University of Rome, I-00185 Rome, Italy
}
\address{
${}^\P$
  INFN - Sezione di Firenze, Polo Scientifico, Via Sansone 1, 
  I-50019, Sesto Fiorentino (FI), Italy 
}
\address{
${}^\dagger$\
Dipartimento di Fisica, Universit\`a di Padova, and INFN, Sezione di Padova, Via Marzolo 8,  I-35131 Padova, Italy}
\address{
  ${}^\ddag$\
  Dipartimento di Fisica, Universit\`a di Lecce, and INFN - Sezione di Lecce,
  Via Arnesano, CP 193, I-73100 Lecce, Italy}

\begin{abstract}
We study the behaviour of spinning test particles in the Schwarzschild spacetime.
Using  Mathisson-Papa\-pe\-trou equations of motion
we confine our attention to spatially circular orbits and search for observable effects
which could eventually discriminate among
the standard supplementary conditions namely the Cori\-nal\-de\-si-Papa\-pe\-trou, Pirani and Tulczyjew.
We find that if the world line chosen for the multipole reduction and whose unit tangent we denote as $U$ is a circular orbit then  also the generalized 
momentum $P$ of the spinning test particle is tangent to a circular orbit even though $P$ and $U$ are not parallel four-vectors. 
These  orbits are  shown to exist because the spin induced  tidal forces provide the required acceleration 
no matter what supplementary condition we select.
Of course, in the limit of a small spin the particle's orbit is close of being a circular geodesic and
the (small) deviation of the angular velocities from the geodesic values can be of an arbitrary sign, corresponding to the
possible  spin-up and spin-down alignment to the $z$-axis. 
When two spinning particles orbit around a gravitating source in opposite directions, 
they make one loop with respect to a given static observer with different arrival times.
This difference is termed clock effect.  
We find that a nonzero gravitomagnetic clock effect appears for oppositely orbiting both spin-up or spin-down  particles even in the Schwarzschild spacetime.
This allows us to establish a formal analogy with the case of (spin-less) geodesics on the equatorial plane of the Kerr spacetime.
This result can be verified experimentally.
\end{abstract}

\pacno{04.20.Cv}

\section{Introduction}

The gravitomagnetic clock-effect is the asymmetry in the arrival time of a pair of intersecting timelike (see \cite{mas1,cohen,bjm} and references therein) or null 
\cite{ciufo1,ciufo2} geodesics, stemming from the same source and having  opposite azimuthal 
angular momentum as measured by a given observer. Hereafter we shall refer the co/counter-rotation as with respect to a fixed sense of variation of the azimuthal angular coordinate. 
In the case of a static observer and of timelike spatially circular geodesics the coordinate time delay is given by
\beq
\Delta t_{(+,-)}=2\pi \left(\frac1{\zeta_+}+\frac1{\zeta_-}\right),
\eeq
where $\zeta_\pm$ denote angular velocities of two opposite rotating geodesics.
In the Schwarzschild solution one has $\zeta_+=-\zeta_-$ and so the clock effect vanishes. 

Here we extend the notion of clock effect to non geodesic circular trajectories considering 
co/counter-rotating spinning-up/spinning-down particles. In this case the time delay is nonzero and can be measured.

The  equations of motion  for a spinning test particle in a given gravitational background were deduced
by Mathisson and Papapetrou \cite{math37,papa51} and read
\begin{eqnarray}
\label{papcoreqs1}
\frac{DP^{\mu}}{\rmd \tau_U}&=&-\frac12R^{\mu}{}_{\nu\alpha\beta}U^{\nu}S^{\alpha\beta}\equiv F^{\rm (spin)}{}^{\mu}\ , \\
\label{papcoreqs2}
\frac{DS^{\mu\nu}}{\rmd \tau_U}&=&P^{\mu}U^{\nu}-P^{\nu}U^{\mu}\ ,
\end{eqnarray}
where $P^{\mu}$ is the total four-momentum of the particle, and $S^{\mu\nu}$ is a (antisymmetric) spin tensor; 
$U$ is the timelike unit tangent vector of the \lq\lq center of mass line'' used to make the multipole reduction. 
The test character of the particle under consideration refers to its mass as well as to its spin, since both quantities should not be 
large enough to contribute to the background metric. In what follows, with the magnitude of the spin of the particle, with the mass and  with a natural lenghtscale associated to the gravitational background we will construct an adimensional parameter as a smallness indicator, which we retain to the first order only so that the test character of the particle  be fully satisfied.  
Moreover, in order to have a closed set of equations Eqs.~(\ref{papcoreqs1})  and (\ref{papcoreqs2}) must be completed 
with  supplementary conditions (SC) whose standard choices in the literature are the
\begin{itemize}
\item[1.]
Corinaldesi-Papapetrou \cite{cori51} conditions (CP): $S^{t\nu}=0$,
\item[2.]
Pirani \cite{pir56} conditions (P): $S^{\mu\nu}U_\nu=0$, 
\item[3.]
Tulczyjew \cite{tulc59} conditions (T): $S^{\mu\nu}P_\nu=0$.
\end{itemize}
Detailed  studies concerning spinning test particles in general relativity are due to Dixon \cite{dixon64,dixon69,dixon70,dixon73,dixon74}, 
Taub \cite{taub64}, Mashhoon \cite{masspin1,masspin2} and Ehlers and Rudolph \cite{ehlers77}.
The need for a supplementary condition whose choice among the three mentioned above is arbitrary, makes the general relativistic description 
of a spinning test particle somehow unsatisfactory. 
Hence,  following the work of Tod, de Felice and Calvani \cite{tod77}, we shall study here the motion of spinning test particles 
on circular orbits in the Schwarzschild spacetime and compare the results obtained by using   different supplementary conditions.   

In this case, both  $U$ and $P$ are linear combinations of Killing vector fields; this property 
leads to a big simplification since all the Frenet-Serret intrinsic quantities of the world line $U$ (curvature and torsions), the magnitude of $P$, 
the algebraic invariant of $S$, namely $s^2=\frac12S_{\mu\nu}S^{\mu\nu}$, and other kinematically relevant quantities, will be (covariantly) constant along $U$.

For certain choices of spin alignment, the \lq\lq spin force", which  couples the particle's spin to the spacetime curvature, is able to maintain 
the orbit; in the absence of spin, instead, only geodesic motion  is allowed, unless other external forces are applied.

Finally, let us note that Eqs.~(\ref{papcoreqs1}) and (\ref{papcoreqs2}) define 
the evolution of $P$ and $S$ only along the world line of  $U$, so a correct interpretation of $U$ is that of being tangent to the  {\it true} 
world-line of the spinning particle. The four-momentum  $P$ and the spin tensor $S$ are  defined as vector fields along the trajectory of $U$.

The paper is organized as follows. In section 2 we specify the properties of timelike spatially circular orbits followed by spinning particles in the Schwarzschild spacetime. In subsections 2.1 to 2.3 we characterize the solution of the equations of motion in terms of the supplementary conditions given by
Corinaldesi and Papapetrou, Pirani and Tulczyjew. In section 3 we deduce the clock effect measured by means of the circular orbits and show how this
observable effect can discriminate among different supplementary conditions. 
We summarize our results in the conclusions and finally in the appendix we extend our analysis to the flat spacetime.
 
In what follows Greek indices run from 0 to 3 while Latin indeces run from 1 to 3; the spacetime metric signature is +2 and geometrized units are used such that both the velocity of light in  vacuum $c$ and the gravitational constant $G$ are set equal to one.
 
\section{Spinning particles in Schwarzschild spacetime}

Let us consider the case of the Schwarzschild spacetime, with the metric written in standard Schwarzschildian coordinates,
\beq\fl\quad 
\label{metric}
\rmd  s^2 = -\left(1-\frac{2M}r\right)\rmd t^2 + \left(1-\frac{2M}r\right)^{-1} \rmd r^2 +r^2 (\rmd \theta^2 +\sin^2 \theta \rmd \phi^2),
\eeq
and introduce an orthonormal frame adapted to the static observers
\beq\fl\quad 
\label{frame}
e_{\hat t}=(1-2M/r)^{-1/2}\partial_t, \,
e_{\hat r}=(1-2M/r)^{1/2}\partial_r, \,
e_{\hat \theta}=\frac{1}{r}\partial_\theta, \,
e_{\hat \phi}=\frac{1}{r\sin \theta}\partial_\phi ,
\eeq
with dual 
\begin{equation}\fl\quad 
\omega^{{\hat t}}=(1-2M/r)^{1/2}\rmd t, \, 
\omega^{{\hat r}}=(1-2M/r)^{-1/2}\rmd r, \, 
\omega^{{\hat \theta}}=r \rmd \theta, \,
\omega^{{\hat \phi}}=r\sin \theta \rmd\phi\ .
\end{equation}
Let us assume that $U$ is  tangent to a (timelike) spatially circular orbit, hereafter denoted as the $U$-orbit, with
\beq
\label{orbita}
U=\Gamma [\partial_t +\zeta \partial_\phi ]=\gamma [e_{\hat t} +\nu e_{\hat \phi}], \qquad \gamma=(1-\nu^2)^{-1/2}\ ,
\eeq
where $\zeta$  is
the angular velocity with respect to infinity and $\Gamma$ is a normalization factor
\beq
\Gamma =\left( -g_{tt}-\zeta^2g_{\phi\phi} \right)^{-1/2}
\eeq
which assures that $U\cdot U=-1$; here dot means scalar product with respect to the metric (\ref{metric}). The angular velocity $\zeta$
is related to the local proper linear velocity $\nu$ measured in the frame (\ref{frame}) by
\beq
\zeta=\sqrt{-\frac{g_{tt}}{g_{\phi\phi}}} \nu .
\eeq
Here $\zeta$ and therefore also $\nu$ are assumed to be constant along the $U$-orbit. We limit our analysis to  the equatorial plane ($\theta=\pi/2$) of the Schwarzschild solution; as a convention, the physical (orthonormal) component along $-\partial_\theta$, perpendicular to the equatorial plane will be referred to as along the positive $z$-axis and will be indicated by $\hat z$, when necessary.

Among the circular orbits particular attention is devoted to the co-rotating $(\zeta_+)$
 and counter-rotating $(\zeta_-)$ timelike circular geodesics, having respectively 
$\zeta_\pm\equiv \pm\zeta_K=\pm (M/r^3)^{1/2}$, so that 
\beq\fl\quad 
\label{Ugeos}
U_\pm=\gamma_K [e_{\hat t} \pm \nu_K e_{\hat \phi}]\ , \qquad \nu_K=\left[\frac{M}{r-2M}\right]^{1/2}, \qquad \gamma_K=\left[\frac{r-2M}{r-3M}\right]^{1/2}\ ,
\eeq
with the timelike condition $\nu_K < 1$ satisfied if $r>3M$.

It is convenient to introduce the Lie relative curvature \cite{idcf1,idcf2} of each orbit 
\beq
k_{\rm (lie)}=-\partial_{\hat r} \ln \sqrt{g_{\phi\phi}}=-\frac1r\left(1-\frac{2M}{r}\right)^{1/2}=-\frac{\zeta_K}{\nu_K}\ ,
\eeq
as well as a Frenet-Serret (FS) intrinsic frame along $U$ \cite{iyer-vish}, defined by  
\beq\fl\quad 
\label{FSframe}
E_{\hat t}=U\ , \qquad
E_{\hat r}=e_{\hat r}\ , \qquad
E_{\hat z}=e_{\hat z}\ , \qquad
E_{\hat \phi}=\gamma[\nu e_{\hat t} +e_{\hat \phi}]\ ,
\eeq
satisfying the following system of evolution equations
\begin{eqnarray}
\label{FSeqs}
\frac{DU}{d\tau_U}&\equiv&a(U)=\kappa E_{\hat r}\ ,\qquad \,\,\, 
\frac{DE_{\hat r}}{d\tau_U}\,=\,\kappa U+\tau_1 E_{\hat \phi}\ ,\nonumber \\
 \nonumber \\
\frac{DE_{\hat \phi }}{d\tau_U}&=&-\tau_1E_{\hat r}\ , \qquad \qquad 
\frac{DE_{\hat z }}{d\tau_U}\,=\,0,
\end{eqnarray}
where 
\beq
\kappa=k_{\rm (lie)}\gamma^2[\nu^2-\nu_K^2]\ , \qquad
\tau_1=-\frac{1}{2\gamma^2} \frac{d\kappa}{d\nu}=-k_{\rm (lie)}\frac{\gamma^2}{\gamma_K^2}\nu\, ; 
\eeq
in this case the second torsion $\tau_2$ is identically zero.
The dual of (\ref{FSframe}) is given by
\beq\fl\quad 
\label{FSframedual}
\Omega^{\hat t}=-U\ , \qquad
\Omega^{\hat r}=\omega^{\hat r}\ , \qquad
\Omega^{\hat z}=\omega^{\hat z}\ , \qquad
\Omega^{\hat \phi}=\gamma[-\nu \omega^{\hat t} +\omega^{\hat \phi}]\ .
\eeq

To study the motion of spinning test particles on circular orbits let us consider first the evolution equation of the spin tensor (\ref{papcoreqs2}). 
In order to most easily satisfy the Riemann force equation we look for solutions for which the frame components of the spin tensor are constant along the orbit.  Were these quantities not constant, precession effects
would surface adding considerably to the mathematics  and enriching the
general behaviour of the spinning body. This however although referring
to a physically more realistic setting  is driven by the basic clock
effect we are discussing here.

By contracting both sides of Eq.~(\ref{papcoreqs2}) with $U_\nu$, one obtains the following expression for the total four-momentum
\begin{equation}
\label{Ps}
P^{\mu}=-(U\cdot P)U^\mu -U_\nu \frac{DS^{\mu\nu}}{\rmd \tau_U}\equiv
mU^\mu +P_s^\mu\ ,
\end{equation}
where $m$ is the particle's {\it bare} mass, namely the mass it would have were it not spinning and  $P_s=U \rightcontract DS/{\rmd \tau_U}$ ($\rightcontract$ denoting the right contraction operation between tensors in index-free form) is a four-vector orthogonal to $U$. 
As a consequence of (\ref{Ps}), Eq.~(\ref{papcoreqs2}) is equivalent to
\begin{equation}
\label{proiet}
P(U)^\mu_{\alpha}P(U)^\nu_{\beta}\frac{DS^{\alpha\beta}}{\rmd \tau_U}=0\ ,
\end{equation}
where $P(U)^\mu_\alpha=\delta^\mu_\alpha+U^\mu U_\alpha$  projects into the local rest space of $U$; it implies
\begin{equation}
\label{spinconds}
S_{\hat t\hat \phi}=0\ , \qquad S_{\hat r\hat \theta}=0\ , \qquad S_{\hat t\hat \theta}+S_{\hat \phi\hat \theta} \frac{\nu }{\nu_K^2}=0\ .
\end{equation}
From Eqs.~(\ref{FSframe}) - (\ref{FSframedual}) it follows that
\beq
\frac{DS}{d\tau_U}=m_s [\Omega^{\hat \phi}\wedge  U]\ ;
\eeq
hence $P_s$ can be written as
\begin{equation}
\label{ps}
P_s=m_s \Omega^{\hat \phi}\ , 
\end{equation}
where
\begin{equation}
\label{msdef}
m_s\equiv||P_s||=\gamma \frac{\zeta_K}{\nu_K}\left[-\nu_K^2 S_{\hat r\hat \phi}+\nu S_{\hat t\hat r}\right]\ .
\end{equation}
From (\ref{Ps}) and (\ref{ps}) and provided $m+\nu m_s\not=0$, 
the total four-momentum $P$ can be written in the form $P=\mu \, U_p$, with
\begin{equation}\fl\quad 
\label{Ptot}
U_p=\gamma_p\, [e_{\hat t}+\nu_p e_{\hat \phi}]\ , \quad \nu_p=\frac{\nu+m_s/m}{1+\nu m_s/m}\ ,\quad \mu=\frac{\gamma}{\gamma_p}(m+\nu m_s)\ ,
\end{equation}
where $\gamma_p=(1-\nu_p^2)^{-1/2}$; $U_p$ is a time-like unit vector, hence $\mu$ has the property of a physical mass.
The first of eqs. (\ref{Ptot}) tells us that as the particle's center of mass moves along the $U$-orbit, the momentum $P$ is istantaneously (namely at each point of the $U$-orbit) parallel to a unit vector
which is tangent to a spatially 
circular orbit, hereafter denoted as $U_p$-orbit, which intersects the $U$-orbit at each of its points.  
Although $U$- and $U_p$-orbits have the same spatial projection into the $U$-quotient space, there exists in the spacetime one $U_p$-orbit for each  point of the $U$-orbit where the two intersect. The $U_p$-orbits then have no direct physical meaning apart from allowing us to identify $\mu$ as the particle's total rest mass in a frame adapted to the $U_p$-orbit. 
Let us now consider the equation of motion (\ref{papcoreqs1}). 
The spin-force is equal to:
\begin{eqnarray}
F^{\rm (spin)}=\gamma \, \zeta_K^2 \left[2S_{\hat t\hat r}+\nu S_{\hat r\hat \phi}\right]e_{\hat r}-\gamma\frac{\nu}{r^2}S_{\hat \theta\hat \phi}e_{\hat \theta}\ ,
\end{eqnarray}
while the term on the left hand side of Eq.~(\ref{papcoreqs1}) can be written, from (\ref{Ps}) and (\ref{ps}),  as 
\beq
\label{motrad}
\frac{DP}{\rmd \tau_U}=m a(U)+m_s \frac{DE_{\hat \phi}}{\rmd \tau_U},
\eeq
where $a(U)=\kappa e_{\hat r}$ and $DE_{\hat \phi}/{\rmd \tau_U}=-\tau_1 e_{\hat r}$ are given in (\ref{FSeqs}), and the quantities  $\mu, m, m_s$ are constant along the world line of $U$. 
The term $a(U)$ is the acceleration of the $U$-orbit and it vanishes if the latter is a geodesic ($\nu=\nu_K$).
The term $DE_{\hat\phi}/d\tau_{_U}=-\tau_1  e_{\hat r}$ represents the first torsion of $U$;  
the term $-m_s\tau_1 e_{\hat r}$ is a spin-rotation coupling force that has been predicted by Mashhoon \cite{mash88} and  has been studied in detail in \cite{bjdf0,bjdf}: it is a sort of centrifugal force (modulo factors), which of course vanishes when the particle is at rest ($\nu=0$).

Since $DP/{\rmd \tau_U}$ is directed radially as from (\ref{FSeqs}) and (\ref{motrad}), Eq.~(\ref{papcoreqs1}) requires that  $S_{\hat \theta\hat \phi}=0$
(and therefore also $S_{\hat t\hat \theta}=0$
from (\ref{spinconds})); hence Eq.~(\ref{papcoreqs1}) can be written as
\beq
m\kappa -m_s\tau_1-F^{\rm (spin)}_{\hat r}=0\ ,
\eeq
or, more explicitly,
\begin{eqnarray}
\label{eqmoto}
0=m\gamma [\nu^2-\nu_K^2]+m_s\frac{\gamma \nu}{\gamma_K^2}+\nu_K\zeta_K\left[2S_{\hat t\hat r}+\nu S_{\hat r\hat \phi}\right]\ . 
\end{eqnarray}

Summarizing, from the equations of motions (\ref{papcoreqs1}) and (\ref{papcoreqs2}) and  before imposing a supplementary condition, the spin tensor turns out  to be completely determined 
by two components only, namely $S_{\hat t\hat r}$ and $S_{\hat t\hat \phi}$,
related by  Eq.~(\ref{eqmoto}). The spin tensor then takes the form
\begin{equation}
S=\omega^{\hat r}\wedge [S_{\hat r\hat t}\omega^{\hat t}+S_{\hat r\hat \phi}\omega^{\hat \phi}]\ .
\end{equation}
It is useful to introduce together with the quadratic invariant 
\beq
s^2=\frac12 S_{\mu\nu}S^{\mu\nu}=-S_{\hat r\hat t }^2+S_{\hat r \hat \phi}^2\ , 
\eeq
another frame adapted to $U_p$ given by
\beq
E^p_0=U_p, \quad E^p_{1}=e_{\hat r}, \quad E^p_2=\gamma_p (\nu_p e_{\hat t}+e_{\hat \phi}), \quad E^p_3=e_{\hat z}\ ,
\eeq
whose 
dual frame is denoted by $\Omega^p{}^{\hat a}$.
 
To discuss the features of the motion we need to supplement Eq.~(\ref{eqmoto}) with further conditions. 
We will do this in the next section following the standard approaches existing in the literature.

\subsection{The Corinaldesi-Papapetrou (CP) supplementary conditions}

The CP supplementary conditions require $S_{\hat t \hat r}=0$, so that 
\beq
\label{SdefCP}
S=s \, \omega^{\hat r}\wedge \omega^{\hat \phi}
\eeq
and Eq.~(\ref{eqmoto}) reduces to
\begin{equation}
\label{eqmotoCP}
0=(\nu^2-\nu_K^2)[1-M{\hat s}\gamma \nu \nu_K\, \zeta_K]\ ,
\end{equation}
where ${\hat s}=\pm |{\hat s}|=\pm |s|/(mM)$ denotes the signed magnitude of the spin per unit (bare) mass $m$ of the test particle and $M$ of the black hole. It is worth noting that the vector dual of $S$ given by Eq.~(\ref{SdefCP}) is directed along the $z$-axis.
The first clear limitation of Eq.~(\ref{eqmotoCP}) is that of being
 meaningful only if $\nu \not =0$.

There arise two cases.
\begin{itemize}
\item[a)]
$\nu\not=\pm\nu_K$.

Eq.~(\ref{eqmotoCP}) gives
\begin{equation}
\label{solCPcase1}
{\hat s}=\frac1{M\gamma \nu\nu_K \zeta_K} .
\end{equation}
The total four momentum (\ref{Ptot}) turns out to be 
\begin{equation}
P=mU-\frac{m}{\nu}E_{\hat \phi}=- mM{\hat s} \nu_K \zeta_K e_{\hat \phi}\ ,
\end{equation}
since $m_s=-m/\nu$, using  (\ref{solCPcase1}). 
However, this solution is manifestly unphysical, since it would imply $P$ is spatial.

\item[b)] 
$\nu=\pm\nu_K$.

Eq.~(\ref{eqmotoCP}) is identically satisfied for all values of ${\hat s}$ showing that
the center of mass line, namely the $U$-orbit is a geodesic even in the presence of a spin force acting radially. 
The latter causes the $U_p$-orbits not to be geodesics. In this case
the total four momentum $P$ is given by (\ref{Ptot}) with
\begin{equation}\fl\quad 
\label{solCPcase2}
\frac{m_s}m=-\gamma_K \nu_K\, \zeta_K M{\hat s}\qquad \hbox {\rm and} \qquad \nu_p=\pm\nu_K \frac{1\mp\gamma_K \zeta_K M{\hat s}}{1\mp\gamma_K \nu_K^2\zeta_K M{\hat s}}\ .
\end{equation}
In the limit of small values of the spin parameter $|{\hat s}|\ll 1$, the linear velocity $\nu_p$ in (\ref{solCPcase2}) reduces to
\begin{equation}
\label{solCPcase2exp}
\nu_p= \pm\nu_K -\frac{\zeta_K\nu_K}{\gamma_K}M{\hat s}+O({\hat s}^2)\ , 
\end{equation}
to first order in ${\hat s}$. The corresponding angular velocity $\zeta_p$ and its reciprocal are 
\begin{equation}
\label{zetaCP}\fl\quad 
\zeta_p= \pm\zeta_K \left[1\mp\frac{\zeta_K}{\gamma_K}M{\hat s}\right]+O({\hat s}^2)\ , \qquad 
\frac1{\zeta_p}= \pm\frac{1}{\zeta_K}+\frac{M{\hat s}}{\gamma_K}+O({\hat s}^2)\ .
\end{equation}
\end{itemize}

\subsection{The Pirani (P) supplementary conditions}

\begin{figure} 
\typeout{*** EPS figure 1}
\begin{center}
\includegraphics[scale=0.45]{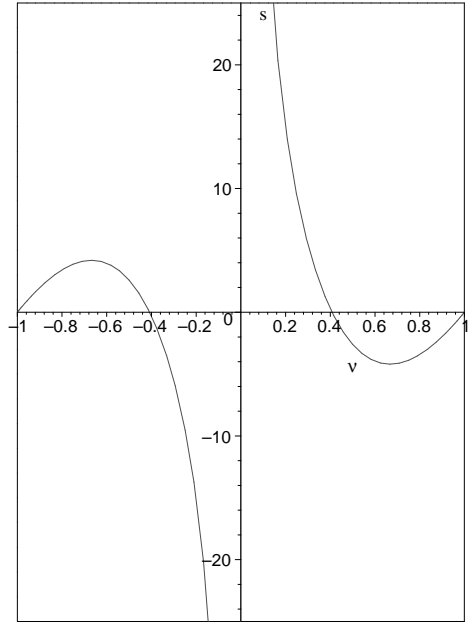}
\end{center}
\caption{In the case of Pirani's supplementary conditions, the spin parameter ${\hat s}$ is plotted as a function 
of the linear velocity $\nu$, for $M=1$ and $r=8$.
From (\ref{solP}) we have that ${\hat s}$ vanishes for $\nu=\pm\nu_K\approx\pm0.408$.
The local maximum and minimum occur at points ($\nu\approx\pm0.668$, ${\hat s}\approx\mp4.2$).
In this plot as well as in the next Fig. 2 large values of $\hat s$ have not a direct physical interpretation in the framework of the Mathisson-Papapetrou model: in such a case in fact the spinning particle loses its test character.
}  
\label{fig:1}
\end{figure}

The P supplementary conditions require $S_{\hat r \hat t}+S_{\hat r \hat \phi}\nu=0$ ($S^{\mu\nu}U_\nu=0$)
or
\beq
S= s \, \omega^{\hat r}\wedge \Omega^{\hat \phi}, \qquad \Omega^{\hat \phi}=\gamma [-\nu \omega^{\hat t}+\omega^{\hat \phi}],
\eeq
so that $(S_{\hat r \hat t},S_{\hat r \hat \phi})=(-s\gamma\nu, s\gamma)$ and
Eq.~(\ref{eqmoto}) reduces to
\begin{equation}
\label{eqmotoP}
0=\nu^2-\nu_K^2+\frac{\nu}{\nu_K}\zeta_K M{\hat s}\left[\frac{\gamma^2}{\gamma_K^2}(\nu^2-\nu_K^2)+3\nu_K^2\right]\ ,
\end{equation}
where, as before, the spin per unit mass has been introduced.
By solving this equation with respect to ${\hat s}$, we obtain
\begin{equation}
\label{solP}
{\hat s}=-\frac{\nu_K}{\nu}\frac1{M\zeta_K}\frac{\nu^2-\nu_K^2}{\left[\frac{\gamma^2}{\gamma_K^2}(\nu^2-\nu_K^2)+3\nu_K^2\right]}\ .
\end{equation}
The behavior of the spin parameter ${\hat s}$ as a function of $\nu$ is shown in Fig. \ref{fig:1}, for a fixed value of the radial coordinate $r$. 
Contrary to the previous case Pirani's supplementary conditions imply that geodesic circular motion is possible only if the spin is zero while 
Eq.~(\ref{eqmotoP}) indicates, as for the previous case, that it is not possible to have a particle at rest ($\nu=0$) with finite spin.

In the limit of small ${\hat s}$, the preceding expression leads to
\begin{equation}
\label{solPexpnu}
\nu= \pm\nu_K-\frac32\zeta_K\nu_K M{\hat s}+O({\hat s}^2)\ .
\end{equation}
The corresponding angular velocity $\zeta$ and its reciprocal are
\begin{equation}\fl\quad 
\label{zetaP}
\zeta= \pm\zeta_K \left[1\mp\frac32\zeta_K M{\hat s}\right]+O({\hat s}^2)\ , \qquad \frac1{\zeta}= \pm\frac{1}{\zeta_K}+\frac32 M{\hat s}+O({\hat s}^2)\ .
\end{equation}
The total four momentum $P$ is given by (\ref{Ptot}) with
\begin{equation}
\frac{m_s}m=\gamma^2\frac{\zeta_K }{\nu_K} M{\hat s}[\nu^2-\nu_K^2]\ , \qquad  \nu_p=\nu+O({\hat s}^2)\ .
\end{equation}
The  angular velocity $\zeta_p$ and its reciprocal are
\begin{equation}
\label{zetapP}
\zeta_p= \zeta+O({\hat s}^2)\ , \qquad \frac1{\zeta_p}= \frac{1}{\zeta}+O({\hat s}^2)\ ,
\end{equation}
with $\zeta$ given by Eq.~(\ref{zetaP}).

\subsection{The Tulczyjew (T) supplementary conditions}

The T supplementary conditions require $S_{\hat r \hat t}+S_{\hat r \hat \phi}\nu_p=0$ ($S^{\mu\nu}P_\nu=0$), 
or
\beq
S= s\, \omega^{\hat r}\wedge \Omega^p{}^{\hat \phi}, \qquad \Omega^p{}^{\hat \phi}=\gamma_p [-\nu_p \omega^{\hat t}+\omega^{\hat \phi}]
\eeq
so that $(S_{\hat r \hat t},S_{\hat r \hat \phi})=(-s\gamma_p\nu_p, s\gamma_p)$ and Eq.~(\ref{eqmoto}) reduces to
\begin{eqnarray}
\label{eqmotoT}
0=\gamma(\nu^2-\nu_K^2)+\gamma_p\frac{\zeta_K}{\nu_K}M{\hat s}\left[\frac{\gamma^2}{\gamma_K^2}\nu(\nu\nu_p-\nu_K^2)+\nu_K^2(2\nu_p+\nu)\right]\ .
\end{eqnarray}
Solving with respect to $\hat s$, we obtain
\begin{eqnarray}
\label{nupTeqs}
{\hat s}&=&-\frac{\nu_K}{M\gamma_p\gamma\zeta_K}\frac{\nu^2-\nu_K^2}{[(1-3\nu_K^2)\nu^2+2\nu_K^2]\nu_p-\nu\nu_K^2(\nu^2-\nu_K^2)}\ .
\end{eqnarray}
Recalling its definition (\ref{msdef}), $m_s$ becomes
\begin{equation}
\frac{m_s}m=\gamma\gamma_p\frac{\zeta_K}{\nu_K} M{\hat s}[\nu\nu_p-\nu_K^2]\ , 
\end{equation}
and using (\ref{Ptot}) for $\nu_p$, we obtain 
\beq
\label{sfromms}
{\hat s}=-\frac{\nu_K}{M\gamma_p\gamma\zeta_K}\frac{\nu-\nu_p}{(1-\nu\nu_p)(\nu\nu_p-\nu_K^2)}\ ; 
\eeq
this condition  must be considered together with (\ref{nupTeqs}).
Relations (\ref{nupTeqs}) and (\ref{sfromms})   imply that the spinless case ($\hat s=0$) is only compatible with $\nu=\nu_p=\pm\nu_K$.
By eliminating ${\hat s}$ from  equations (\ref{sfromms}) and (\ref{nupTeqs}) and solving with respect to $\nu_p$ we have that 
\begin{equation}
\label{nupsol}
\nu_p^{(\pm)}=\frac12\frac{\nu_K}{\nu^2+2\nu_K^2}\{3\nu\nu_K\pm[\nu^2(13\nu_K^2+4\nu^2)-8\nu_K^4]^{1/2}\}\ .
\end{equation}
Since the case ${\hat s} \ll 1$ is the only physically relevant case we shall consider then of the two branches of the solution (\ref{nupsol})
we have to take only $\nu_p^{(+)}$ when $\nu>0$ and $\nu_p^{(-)}$ when $\nu<0$.
By substituting $\nu_p=\nu_p^{(\pm)}$ for instance into Eq.~(\ref{nupTeqs}), we obtain a relation between $\nu$ and ${\hat s}$.
The reality condition of (\ref{nupsol}) requires that $\nu$ takes values outside the interval $({\bar \nu}_-,{\bar \nu}_+)$, 
with ${\bar \nu}_{\pm}=\pm\nu_K\sqrt{2}\sqrt{-13+3\sqrt{33}}/4\simeq \pm 0.727 \nu_K$; moreover, 
the timelike condition for $|\nu_p| <1$ is satisfied for all values of $\nu$ outside the same interval.

\begin{figure} 
\typeout{*** EPS figure 2}
\begin{center}
\includegraphics[scale=0.45]{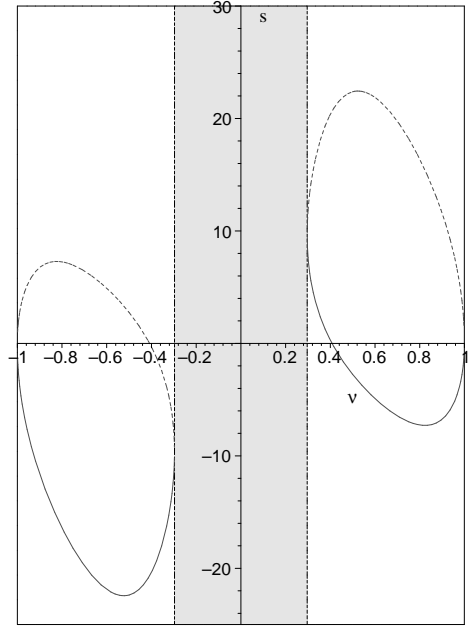}
\end{center}
\caption{
In the case of Tulczyjew's supplementary conditions, the spin parameter ${\hat s}$ is plotted  as a function of the linear velocity 
$\nu$ for $M=1$ and $r=8$ (and so ${\bar \nu}_{\pm}\approx\pm0.297$; the shaded region contains the forbidden values of $\nu$).
At the boundary of the forbidden region the spin is finite ($\hat s \approx \pm 9.7$).
Note that ${\hat s}$ vanishes for $\nu=\pm\nu_K\approx\pm0.408$ (and so $\nu_p^{(\pm)}=\pm\nu_K$ too)
and $\hat s(\bar \nu_{\pm})=\pm9.683$;  
The plot have two branches corresponding to $\nu=\nu_p^{(+)}$ (solid) and to $\nu=\nu_p^{(-)}$ (dashed).
}  
\label{fig:2}
\end{figure}

The behavior of the spin parameter ${\hat s}$ as a function of $\nu$ is shown in Fig. \ref{fig:2}, for a fixed value of the radial coordinate $r$. 
This plot shows that there exists a range of velocities $\nu$ which is
forbidden for physical $U$-orbits.
A natural explanation of this is the lack of centrifugal forces strong enough to balance the spin force.

A linear relation between $\nu$ and ${\hat s}$ can be obtained in the limit of small ${\hat s}$:
\begin{equation}
\label{solTexpnu}
\nu= \pm\nu_K-\frac32\zeta_K\nu_K M{\hat s}+O({\hat s}^2)\ .
\end{equation}
From this approximate solution for $\nu$ we also have that 
\begin{equation}
\nu_p^{(\pm)}= \pm\nu_K-\frac32\zeta_K\nu_K M{\hat s}+O({\hat s}^2)\ ,
\end{equation}
and the total four momentum $P$ is given by (\ref{Ptot}) with 
\begin{equation}
\nu_p=\nu+O({\hat s}^2)\ .
\end{equation}
We want to stress again that $\nu$ is equal to $\nu_p$ only up to first order in ${\hat s}$.
The angular velocities $\zeta$, $\zeta_p$ and their reciprocals coincide with the corresponding ones derived in the case of P supplementary conditions (see (\ref{zetaP}) and (\ref{zetapP}) respectively).

\section{Clock-effect for spinning test particles}

As we have  seen in all cases examined above, spinning test particles move on circular orbits which, to first order in the spin parameter $\hat s$, 
are close to a geodesic (as expected), with
\beq
\label{clock}
\frac{1}{\zeta_{(SC,\pm,\pm)}}=\pm \frac{1}{\zeta_K} \pm M|{\hat s}| {\mathcal J}_{SC}\ ,
\eeq 
where ${\mathcal J}_{CP}=0$ and ${\mathcal J}_P={\mathcal J}_T=3/2$.
Eq.~(\ref{clock}) identifies these orbits according to the chosen supplementary condition, the signs in front of $1/\zeta_K$ corresponding to co/counter-rotating 
orbits while the signs in front of ${\hat s}$ refer to a positive or negative spin direction along the $z$-axis; for instance, the quantity 
$\zeta_{(P,+,-)}$ denotes the angular velocity of $U$, 
derived under the choice of Pirani's supplementary conditions and corresponding to a co-rotating orbit $(+)$ with spin-down $(-)$ alignment, etc.
Therefore one can measure the difference in the arrival times after one complete revolution with respect to a 
static observer. The coordinate time difference in given by: 
\beq\label{deltat}
\Delta t_{(+,+;-,+)}= 2\pi \left(\frac{1}{\zeta_{(SC,+,+)}}+\frac{1}{\zeta_{(SC,-,+)}}\right)=4 \pi M|{\hat s}| {\mathcal J}_{SC}
\eeq
and analogously for $\Delta t_{(+,-;-,-)}$. A similar result can be obtained referring to any circularly rotating observer, with a slight 
modification of the discussion as can be found in the literature \cite{bjm}.
Eq.~(\ref{deltat}) states that the measurement of a finite 
 difference in the arrival times,  proportional to the (small) spin per unit mass of the particle would exclude the CP supplementary condition 
as the one selected by nature because that would imply no clock effect, while to first order in $\hat s$ one cannot distinguish between  
Pirani and Tulczyjew's  supplementary conditions. Moreover this effect creates an interesting parallelism with the clock-effect in  Kerr spacetime. 
In the case discussed here, in fact, it is the spin of the particle which 
creates a nonzero clock effect while in the Kerr metric this effect is induced by the {\it rotation} of the spacetime even  
to geodesic spinless test particles. 
The latter have angular velocities
\beq
\frac1{\zeta_{({\rm Kerr})\,\pm}}= \pm\frac{1}{\zeta_K}+a,
\eeq
where $a$ is the angular momentum per unit mass of the Kerr black hole, hence
 Kerr's clock effect reads:
\beq
\Delta t_{(+,-)}=2\pi \left(\frac1{\zeta_{({\rm Kerr})\,+}}+\frac1{\zeta_{({\rm Kerr})\,-}}\right)=4\pi a\ ,
\eeq
This complementarity suggests a sort of equivalence principle: to first order in the spin, a static observer cannot decide whether it measures 
a time delay of spinning clocks in a non rotating spacetime or a time delay of  non spinning  clocks moving on geodesics in a rotating spacetime.

\section{Conclusions}

Spinning test particles in  circular motion around a Schwarzschild black hole have been discussed in the framework of the 
Mathisson-Papapetrou approach supplemented by standard conditions.
In the limit of small spin, the orbit of the particle is close to a circular geodesic and
the difference in the angular velocities with respect to the geodesic value can be of arbitrary sign, 
corresponding to the two spin-up and spin-down orientations along the $z$-axis. 
For co-rotating and counter-rotating both spin-up (or both spin-down) test particles a nonzero gravitomagnetic clock effect 
appears even in the Schwarzschild spacetime, allowing a formal analogy with the case of (spin-less) geodesics on 
the equatorial plane of the Kerr spacetime. The observable effects can be used to discriminate among the different supplementary conditions.

\section*{Acknowledgments}

We are grateful to Prof. B. Mashhoon and Prof. R.T. Jantzen for a critical reading of the manuscript.
Part of this work was done when one of us (F. de F.) was at the Instituto Venezolano de Investigaciones Cientificas (IVIC) nearby Caracas. The director of that institution is thanked for his warm hospitality and the Consiglio Nazionale delle Ricerche (CNR) of Italy is thanked for support.

\appendix
\section{Spinning particles in flat spacetime}

Let us consider a spinning test particle moving along a circular orbit in a Minkowski spacetime, described by the metric
\beq
\rmd  s^2 = -\rmd t^2 + \rmd r^2 + \rmd z^2 +  r^2\rmd \phi^2
\eeq
in cylindrical coordinates ($t, r, z, \phi$), and introduce the orthonormal frame
\beq
e_{\hat t}=\partial_t\ , \qquad
e_{\hat r}=\partial_r\ , \qquad
e_{\hat z}=\partial_z\ , \qquad
e_{\hat \phi}=\frac{1}{r}\partial_\phi .
\eeq
The 4-velocity $U$ associated to a generic timelike circular orbit in this metric is given by
\beq
\label{Uflat}
U=\Gamma [\partial_t +\zeta \partial_\phi ]=\gamma [e_{\hat t} +\nu e_{\hat \phi}], \qquad \gamma=(1-\nu^2)^{-1/2}\ ,
\eeq
where 
$$
\Gamma =\left( 1-r^2\zeta^2 \right)^{-1/2}\ , \qquad \zeta=\frac{\nu}{r}\ .
$$
It is useful to set also in this case along each circular orbit a Frenet-Serret frame $\{E_i\}$ as in (\ref{FSframe}), with
\beq
\kappa=-\gamma^2\frac{\nu^2}{r}\ , \qquad
\tau_1=\gamma^2\frac{\nu}{r}\ . 
\eeq

The Mathisson-Papapetrou equations of motion (\ref{papcoreqs1}) and (\ref{papcoreqs2}) reduce to
\begin{eqnarray}
\label{flatspineqs1}
\frac{DP}{\rmd \tau_U}&=&0\ ,\\ 
\label{flatspineqs2}
\frac{DS}{\rmd \tau_U}&=&P\wedge U\ .
\end{eqnarray}
In order to handle with the various supplementary conditions simultaneously it is convenient to write the spin tensor $S$ in the form  
\beq
\label{Sdefflat}
S=s\, {\tilde E_{\hat r}}\wedge {\tilde E_{\hat \phi}}\ ,
\eeq
being
$$
\tilde E_{\hat r}\equiv E_{\hat r}\ , \qquad
\tilde E_{\hat z}\equiv E_{\hat z}\ , \qquad
\tilde E_{\hat \phi}=\tilde\gamma[U+\tilde\nu E_{\hat \phi}]\ ,
$$
a spatial frame adapted to an observer  with 4-velocity
\beq
\tilde U=\tilde\gamma [U+\tilde\nu E_{\hat \phi}]\ ,
\eeq
so that $S=s\, {}^{*}[{\tilde U}\wedge {\tilde E_{\hat z}}]\ $
and  the associated spin vector results to be directed along the $z$-axis.
The total four momentum $P$ is given by (\ref{Ptot}) with
\begin{equation}
\label{msflat}
m_s=s\tau_1\tilde\gamma[\nu+\tilde\nu]\ ,
\end{equation}
as can be easily shown by means of Eq.~(\ref{flatspineqs2}), by assuming $s$ to be constant along $U$. 
The further condition 
\beq
\label{msflat2}
0=m\nu+m_s\ 
\eeq
arises from Eq.~(\ref{flatspineqs1}). Now, by combining eqs. (\ref{msflat}) and (\ref{msflat2}), we obtain the 
following relation between the quantities ${\hat s}=\pm |{\hat s}|=\pm |s|/m^2$ (the adimensional signed spin magnitude in this case), $\nu$ and $\tilde \nu$: 
\beq
\label{ssolflat}
-m{\hat s}\frac{\gamma^2\tilde\gamma}{r}[\nu+\tilde\nu]=1\ .
\eeq
Finally, this equation have to be supplemented by a SC.   

\begin{itemize}
\item[1)]
CP supplementary conditions:

These conditions require $\tilde\nu=-\nu$, and so $m_s=0$, from Eq.~(\ref{msflat}). 
But Eq.~(\ref{msflat2}) would imply $m\nu=0$, impossible to be satisfied if $m\not=0$. 

\item[2)]
P supplementary conditions:

These conditions require $\tilde\nu=0$; Eq.~(\ref{ssolflat}) becomes
\beq
{\hat s}=-\frac{r}{m\gamma^2\nu}\ .
\eeq
By solving this equation with respect to $\nu$, we obtain
\beq
\nu=\pm 1-\frac{m}{r}\frac{\hat s}{2}+O({\hat s}^2)\ , \qquad 
\frac{1}{\zeta}=\pm r+m\frac{\hat s}{2}+O({\hat s}^2)\ .
\eeq

\item[3)]
T supplementary conditions:

These conditions require $\tilde\nu=-(\nu-\nu_p)/(1-\nu\nu_p)$. 
By using Eq.~(\ref{msflat2}) and the definition (\ref{Ptot}) of $\nu_p$, we have that $\nu_p=0$, and so $\tilde\nu=-\nu$.
Therefore, the situation is analogous to the case 1) of CP supplementary conditions.
\end{itemize}

So apparently in a flat spacetime a spinning particle can be set in motion on a circular orbit only under Pirani's supplementary conditions.

\end{document}